\documentclass[twocolumn,showpacs,preprintnumbers,amsmath,amssymb]{revtex4}
\usepackage{graphicx}
\usepackage{dcolumn}
\usepackage{bm}
\begin{document}
\newcommand{\etal}{{\it{et~al}.}}

\title{Relativistic plasma control for single attosecond pulse generation}

\author{T.~Baeva$^{1,*}$, S.~Gordienko$^{1,2}$ and A.~Pukhov$^1$}

\affiliation{$^1$Institut f\"ur Theoretische Physik I,
             Heinrich-Heine-Universit{\"a}t D\"usseldorf, D-40225, Germany \\
             $^2$L.~D.~Landau Institute for Theoretical Physics, Moscow, Russia \\
             $^*$Electronic address: tbaeva@thphy.uni-duesseldorf.de
            }

\date{\today}

\begin{abstract}
{\noindent 
  To describe the high harmonic generation at plasma
  surfaces in the relativistic regime, we introduce the concept of {\it
  apparent reflection point} (ARP). It appears to an external observer
  that the radiation electric field is zero at the ARP. The relativistic
  dynamics of the ARP completely defines the generation of high
  harmonics and attosecond pulses. The ARP velocity is a
  smooth  function of time. The corresponding $\gamma$-factor,
  however, has sharp spikes at the times when the tangential vector
  potential vanishes and the surface velocity becomes close to the speed of
  light. We show that managing the laser polarization, one can efficiently
  control the ARP dynamics, e.g., to gate a single (sub-)attosecond
  pulse out of the short pulse train generated by a multi-cycle
  driver. This   relativistic control is demonstrated numerically by
  particle-in-cell simulations.
}   
\end{abstract}


\maketitle

 
  High harmonics generated at plasma surfaces in the relativistic
  regime \cite{Experiment,Lichters1996,Gordienko2004}
  are a promising new source of short wavelength radiation and
  attosecond pulses, necessary for the study of ultra-fast processes in atoms, 
  molecules and solids at intensities significantly higher than those obtained 
  from strong field laser-atom interactions \cite{Krausz,Corkum}. 

  The plasma harmonics are generated in the relativistic regime, when
  the laser pulse intensity $I \gg 10^{18}$~W/cm$^2$. Laser pulses
  in this intensity range usually are several cycles long. As a consequence,
  the reflected radiation contains a comb of attosecond pulses. 
  Yet, applications like molecular imaging \cite{Corkum,Lein} or
  quantum control \cite{QuantumControl} usually require a
  single short pulse to prevent undesirable effects, e.g., Coulomb
  explosion. The single attosecond pulse can be selected either
  using a phase-stabilized single cycle laser \cite{Krausz} or 
  controlling the atomic response by time-dependent laser
  polarization \cite{Misha,brixner}. 

  It was shown recently by particle-in-cell (PIC) simulations that in the
  $\lambda^3-$regime, when a single-cycle laser pulse is focused down to 
  a spot of a one wavelength size, a single attosecond pulse running
  in a particular direction can be isolated \cite{naumova}. However,
  in the case of a more realistic several-cycle laser pulse, such
  casual separation can be difficult.

\begin{figure}
    \centerline{\includegraphics[width=8cm,clip]{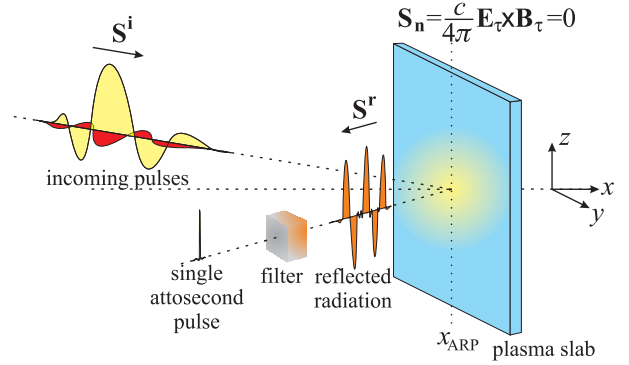}}
    \caption{Geometry for relativistic plasma control of attosecond
    plasma surface dynamics. The high intensity driving pulse is
    polarization-managed through a low intensity controlling
    pulse. After proper filtering of the reflected radiation, a
    single attosecond pulse can be isolated. The interaction process
    can be described in terms of the apparent reflection point
    (ARP). It appears to an external observer that the radiation
    electric field and the Poynting vector $\bf S$ are zero at the ARP.} 
    \label{fig:sketch}
  \end{figure}

  In the present work we show for the first time that the managed
  time-dependent polarization (from elliptical to linear and back) of
  the incident laser pulse allows to gate a single (sub-)attosecond
  pulse from relativistically driven plasma surface in a well
  controlled way. Physically, the laser pulse 
  polarization controls the relativistic $\gamma-$factor of the apparent
  reflection point (ARP) as seen by the observer. Although the driving
  laser pulse can be long and intense, the ARP velocity becomes highly
  relativistic only at the specific time, when the vector
  potential component tangential to the plasma surface vanishes. Exactly
  at this moment, the single attosecond pulse is emitted. 
  The time-dependent polarization corresponds to two perpendicularly
  polarized laser pulses with slightly different frequencies and a
  well chosen phase shift, see Fig.~\ref{fig:sketch}. Experimentally,
  the time-dependent laser pulse ellipticity can be achieved by the
  femtosecond polarization pulse shaping techniques \cite{brixner}.
    
  We consider the interaction of an ultra-intense short laser pulse
  with a slab of overdense plasma. We suppose that the plasma ions are 
  immobile during the short interaction time and study the electron 
  fluid oscillations only. It was shown in \cite{Gordienko2004} that to
  describe the high harmonic generation analytically, the boundary
  condition ${\bf E}_{\tau}=0$ must be used. 
  Here ${\bf E}_{\tau}$ is the electric field component tangential to
  the plasma surface. The physical meaning of this boundary condition
  is quite clear. Let us consider the Poynting vector ${\bf S}=c{\bf
  E}\times {\bf B}/4\pi$ that  
  gives the electromagnetic energy flux. In the vacuum region in front
  of the plasma, it consists of the two parts: the incident flux ${\bf S^i} =
  c{\bf E^i}\times {\bf B^i}/4\pi$ and the reflected one ${\bf S^r} =
  c{\bf E^r}\times {\bf B^r}/4\pi$. 
  Because the plasma slab is overdense, ${\bf S} = 0$ behind the plasma. 
  If we also neglect the small absorption, then the observer
  sees the apparent reflection at the point $x_\text{\tiny ARP}(t)$,
  where the normal component of the Poynting vector ${\bf S_n} = 
  {c\bf E_\tau \times B_\tau/4\pi} = 0$, implied by ${\bf
  E_\tau}(x_\text{\tiny ARP})=0$. Thus the boundary condition ensures the
  energy conservation. The detailed microscopic derivation of
  the boundary condition  ${\bf E_\tau}(x_\text{\tiny ARP})=0$ and its
  relation to the electron dynamics can be found in \cite{TPBdiploma}.
  
 
  The incident laser field in vacuum runs in the positive $x-$direction,
  ${\bf E^i}(x,t)={\bf E^i}(x-ct)$, while the reflected 
  field is translated backwards:  ${\bf E^r}(x,t)={\bf E^r}(x+ct)$. The
  tangential components of these fields interfere  
  destructively at the ARP position $x_\text{\tiny ARP}(t)$, so that
  the implicit equation for the apparent reflection point 
  $x_\text{\tiny ARP}(t)$ is:

  \begin{equation} \label{ARP}
    {\bf E_{\tau}^i}(x_\text{\tiny ARP} - ct) +  {\bf
    E_{\tau}^r}(x_\text{\tiny ARP} + ct) =0. 
  \end{equation}

  \noindent 
  We stress that Eq.~(\ref{ARP}) contains the electromagnetic fields in 
  vacuum. That is why the reflection point $x_\text{\tiny ARP}$ is 
  {\it apparent}. The real interaction within the plasma skin layer can 
  be very complex. Yet, an external observer, who has information about 
  the radiation in vacuum only, sees that ${\bf E_\tau} = 0$ at  $x_\text{\tiny ARP}$.
  Strictly speaking, one cannot ascribe $x_\text{\tiny ARP}$ to any 
  particular density level, e.g., the "critical density". One can only state 
  that the ARP is located within the skin layer at the electron fluid surface.   
  
  The ARP dynamics completely defines the high harmonic generation. In the 
  ultra-relativistic regime, when the dimensionless vector potential 
  $a_0=eA_0/mc^2$ of the laser is large, $a_0^2\gg 1$, we can apply 
  ultra-relativistic similarity theory \cite{Similarity} to characterize this 
  motion. The basic statement of this theory is that when we change
  the plasma density  $N_e$ and the laser amplitude $a_0$
  simultaneously keeping the similarity parameter $S= N_e/a_0 N_c$ constant 
%
%
  the laser-plasma dynamics remains similar. Here $N_c=\omega_0^2 m/4\pi e^2$ 
  is the critical plasma density for the laser pulse with the
  frequency $\omega_0$.
  This means that for different interactions with the same similarity
  parameter $S$=const, the plasma electrons move along the same
  trajectories, while their momenta $\bf p$ scale with the laser
  amplitude: ${\bf p} \propto a_0$. Consequently, the electron momentum
  components tangential and normal to the plasma surface scale
  simultaneously with $a_0$: $\textbf{p}_{\tau} \propto a_0$ and
  $\textbf{p}_{n}\propto a_0$. Therefore, in general, the total
  electron momenta are not perpendicular to the surface. Moreover, the
  characteristic angle between their direction and the surface normal 
  does not depend on $a_0$ provided that $S$ is fixed. 
  
  
  This result is crucial for the plasma surface dynamics. Since we
  consider an ultra-relativistic laser pulse, $a_0^2 \gg 1$, the
  electrons in  the skin layer move with ultra-relativistic
  velocities almost all the time: 
  
  \begin{equation} \label{ElectronVelocity}
    v=c\sqrt{\frac{\textbf{p}_{n}^2+\textbf{p}_{\tau}^2}{m_e^2c^2+     
    \textbf{p}_{n}^2+\textbf{p}_{\tau}^2}}=c(1-O(a_0^{-2})).
  \end{equation}
  
  Yet the relativistic $\gamma$-factor of the plasma surface
  $\gamma_s(t)$ and its velocity $v_s(t)$ 
  behave in a quite different
  way. Let us consider electrons at the very boundary of the
  plasma. The similarity theory states that the electron 
  momenta can be represented as $\textbf{p}_{n}(t)=a_0\textbf{P}_{n}(S,\omega t)$
  and $\textbf{p}_{\tau}(t)=a_0\textbf{P}_{\tau}(S,\omega t)$, 
%
%
  where $\textbf{P}_{n}$ and $\textbf{P}_{\tau}$ are universal functions,
  which do not depend on $a_0$ explicitly, but rather on the similarity 
  parameter $S$ and on the pulse shape. Thus, the plasma surface
  velocity $\beta_s(t)=v_s(t)/c$ and $\gamma_s(t)$ are

  \begin{eqnarray} \label{BoundaryDynamics}      
    \beta_s(t)&=&\frac{p_{n}(t)}{\sqrt{m_e^2c^2+\textbf{p}_{n}^2(t)
    +\textbf{p}_{\tau}^2(t)}} \nonumber\\
    &=&\frac{P_{n}(t)}{\sqrt{\textbf{P}_{n}^2(t)+
    \textbf{P}_{\tau}^2(t)}} - O(a_0^{-2}),\\
    \label{GammaBoundaryDynamics}
    \gamma_s(t)&=&\frac{1}{\sqrt{1-\beta_s^2(t)}}
    =\sqrt{1+\frac{\textbf{P}_{n}^2(t)}{\textbf{P}_{\tau}^2(t)}}+O(a_0^{-2}).
  \end{eqnarray} 
  
  \noindent
  It follows from (\ref{BoundaryDynamics})-(\ref{GammaBoundaryDynamics})
  that the relativistic $\gamma$-factor of the plasma boundary is of the
  order of unity for almost all times, except for those times $t_g$, when 
  the tangential momentum component vanishes 
  $\textbf{p}_{\tau}(S,t_g)=0$.
%
%
  Exactly at these times, there are spikes of the $\gamma$-factor: 
  
  \begin{equation} \label{GammaJumps} 
    \gamma_s=\frac{1}{\sqrt{1-\beta_s^2}}=
    \sqrt{\frac{\textbf{p}_{n}^2+m_e^2c^2}{m_e^2c^2}}\propto a_0.
  \end{equation}
  
  \noindent
  On the contrary, the plasma surface velocity $v_s$ as given
  by Eq.~(\ref{BoundaryDynamics}) is a smooth function. It approaches
  $\pm c$ when the electron momentum parallel to the surface  
  vanishes, i.e., at the same times $t_g$ corresponding to the
  $\gamma$-spikes. The high harmonics are generated at the spikes,
  when the surface velocity is negative and close to $-c$.
  Rigorous microscopic analysis \cite{TPBdiploma} confirms that
  high harmonics are generated by bunches of fast electrons moving 
  towards the laser pulse.
  
  \begin{figure} [h]
    \centerline{\includegraphics[width=7cm,clip]{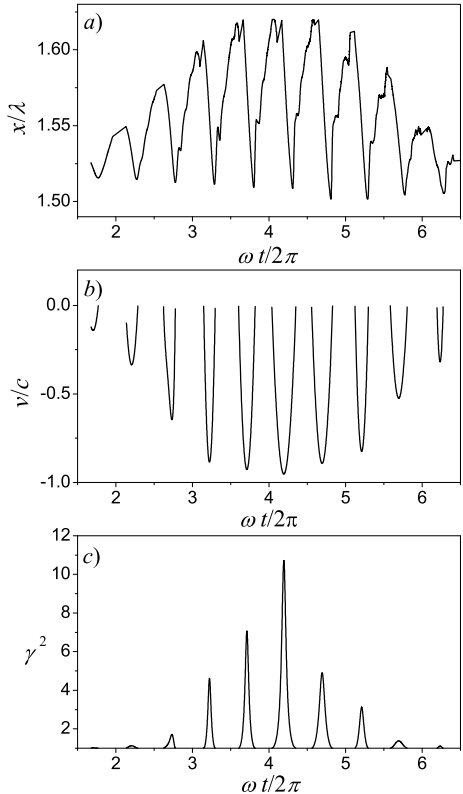}}
    \caption{1D PIC simulation results for the parameters $a_0=20$ and
    $N_e=90 N_c$. a) Oscillatory motion of the point $x_\text{\tiny
    ARP}(t)$ where ${\bf E}_\tau(x(t))=0$.  b) Velocity
    $v_\text{\tiny ARP}(t) = dx_\text{\tiny ARP}(t)/dt$; only the
    negative velocities are shown. Notice that the ARP velocity is a
  smooth function. c) The corresponding $\gamma-$factor
    $\gamma_\text{\tiny  ARP}(t) = 1/\sqrt{1-v_\text{\tiny
    ARP}(t)^2/c^2}$ contains sharp spikes, which coincide
  with the velocity extrema.}
    \label{a20x_v_gamma}
  \end{figure}
  
  It is possible to estimate the width of a $\gamma$-spike. Since the
  surface velocity is a smooth function, we can expand it in Taylor series
  around the maximum as $v_s(t)\approx v_{max}-\alpha\omega_0^2(t-t_{max})^2$,
  where the parameter $\alpha$ depends only on $S$.  
  The width $\Delta t=|t-t_{max}|$ of the $\gamma_s$ spikes is then
  
  \begin{equation} \label{WidthGamma}
    \Delta t\propto\sqrt{1-v^2_{max}/c^2}/(\omega_0\sqrt{\alpha})=1/(\omega_0\sqrt{\alpha}\gamma_{max}).
  \end{equation}
   
  \begin{figure} [h]
    \centerline{\includegraphics[width=7cm,clip]{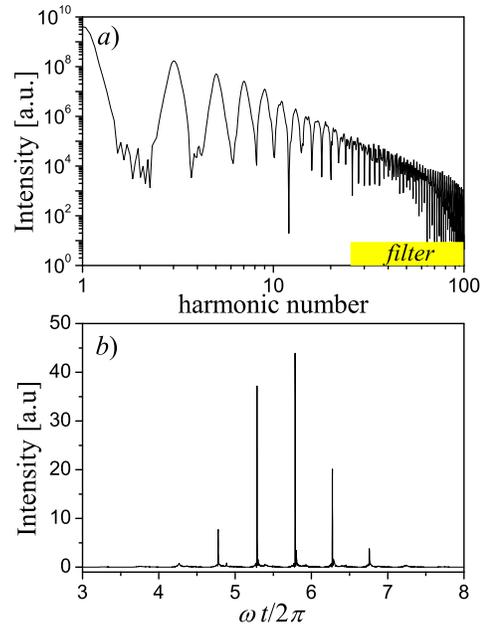}}
    \caption{a) The high harmonics spectrum; b) the train of
    attosecond pulses is obtained after the proper filtering.}
    \label{spectrum_atto}
  \end{figure}
  
  \begin{figure} [h]
    \centerline{\includegraphics[width=7cm,clip]{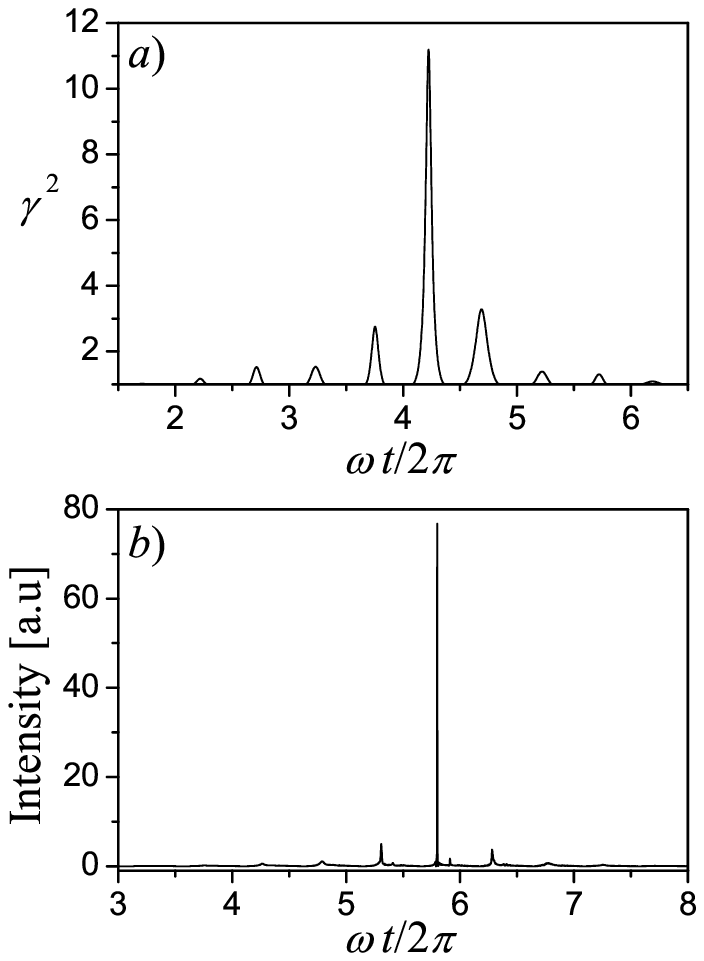}}
    \caption{Generation of a single attosecond pulse using the
    relativistic plasma control. The driver signal has $a_0=20$
             and frequency $\omega_0=1$. The controlling signal has
    $a_c=6$ and frequency $\omega_c=1.25$. The phase difference is
    $\Delta\phi=\pi/8$.  a) the $\gamma$-spikes of the oscillating
    ARP, b) the single attosecond pulse selected via the relativistic
    plasma control.}
    \label{SinglePulse}
  \end{figure}
  
  \noindent
  It follows from (\ref{WidthGamma}) that the $\gamma_s$ spikes get
  higher and narrower when we increase $a_0$, keeping $S=$~const. Thus,
  the spikes behave as quasi-singularities. These ultra-relativistic spikes 
  are the inherent cause for the high harmonic generation. The
  (sub-)attosecond pulses \cite{Gordienko2004} are emitted exactly at the
  times of the spikes.
  
  We study the motion of the plasma boundary and the specific behavior
  of $v_\text{\tiny ARP}$ and $\gamma_\text{\tiny ARP}$ numerically using 
  the 1D PIC code VLPL \cite{vlpl}.
  The plasma slab is initially positioned between $x_L=1.5\lambda$
  and $x_R=3.9\lambda$, where $\lambda = 2\pi/\omega_0$ is the laser
  wavelength. The laser pulse has the gaussian envelope: $a(x,t=0) = a_0
  \exp(-x^2/\sigma^2)\cos(2\pi x/\lambda)$ with $\sigma = 2\lambda$.

  At every time step, the incident and the reflected fields
  are recorded at $x=0$. Being solutions of the wave equation 
  in vacuum, these fields can be easily chased to arbitrary $x$ and $t$.
  To find the ARP position $x_\text{\tiny ARP}$, we solve numerically
  the equation (\ref{ARP}). The trajectory of $x_\text{\tiny ARP}(t)$ 
  for the simulation parameters $a_0=20$ and $N_e/N_c=90$ ($S=4.5$) is 
  presented in Fig. \ref{a20x_v_gamma} a). One can clearly see the oscillatory 
  motion of the point $x_\text{\tiny ARP}(t)$. The equilibrium position 
  is displaced from the initial plasma boundary position $x_L$ due to the 
  mean laser light pressure.

  Since only the ARP motion towards the laser pulse is of importance for 
  the high harmonic generation, we cut out the positive ARP velocities 
  $v_\text{\tiny ARP}(t) = dx_\text{\tiny ARP}(t)/dt$ and calculate only the 
  negative ones, Fig. \ref{a20x_v_gamma} b). The corresponding
  $\gamma$-factor  
  $\gamma_\text{\tiny ARP}(t)=1/\sqrt{1-v_\text{\tiny ARP}(t)^2/c^2}$ is 
  presented in Fig. \ref{a20x_v_gamma} c). Notice that the ARP velocity is a
  smooth function. At the same time, the $\gamma$-factor 
  $\gamma_\text{\tiny ARP}(t)$ contains sharp spikes, which coincide
  with the velocity extrema. These spikes of the surface $\gamma$-factor
  are responsible for the high harmonic generation. 

  The numerically obtained spectrum of the high-harmonics is shown in
  Fig.~\ref{spectrum_atto} a). Filtering out the lower harmonics and
  keeping only the harmonics with $\omega > 25 \omega_0$, we obtain a
  train of short pulses in the reflected radiation,
  Fig. \ref{spectrum_atto} b). 

  For various applications, such as molecule imaging, it is of great
  importance to have a single short pulse, instead of a train of
  short pulses. We have shown above that the attosecond pulses are
  emitted when the tangential components of the surface electron momentum
  vanish. This property can be used to control the high harmonic
  generation and to gate a particular attosecond pulse out of the train. 
 
  In the 1D geometry, the transverse generalized momentum is conserved: 
  $\textbf{p}_{\tau}=e\textbf{A}_{\tau}/c$, where $\textbf{p}_{\tau}$ and
  $\textbf{A}_{\tau}$ are the tangential components of the electron
  momentum $\textbf{p}$ and the vector potential $\textbf{A}$.
  Consequently, the attosecond pulses are emitted when the vector
  potential is zero. If the vector potential vanishes at several
  moments, there are several $\gamma$-spikes and correspondingly,
  after proper filtering, several short pulses are observed in 
  the reflected radiation, see Fig. \ref{spectrum_atto} b).
 
  To select a single attosecond pulse, we must ensure that the vector
  potential $\textbf{A}_{\tau}$ turns zero exactly once. Since
  $\textbf{A}_{\tau}$ has two components, how often it vanishes
  depends on its polarization. For linear polarization it
  vanishes twice per laser period, while for circular or elliptic
  polarization it never equals zero. A laser pulse with a
  time-dependent polarization can 
  be prepared in such a way that its vector potential turns zero just
  once.  The time-dependent polarization corresponds to two
  perpendicularly polarized laser pulses with slightly different
  frequencies, see Fig.~\ref{fig:sketch}.
  Our PIC simulations suggest that a signal with a few percent of the
  driver intensity is sufficient to control the high harmonic
  generation, if the phase difference between the two laser pulses is
  chosen carefully.
  
 
  To demonstrate the relativistic plasma control, we perform a PIC
  simulation where we add the $z$-polarized controlling pulse with 
  amplitude $a_c=6$ and frequency $\omega_c=1.25$ and retain the same
  driver pulse with amplitude $a_0=20$,  frequency $\omega_0=1$ and
  $y$-polarization. We keep the same plasma density $N_e/N_c=90$. The optimal 
  phase difference between the two lasers is found empirically to be
  $\Delta\phi=\pi/8$. The simulation results are presented 
  in Fig.~\ref{SinglePulse} a) and b). Comparing the surface $\gamma-$factor 
  dynamics in the regime of linear polarization,
  Fig.~\ref{a20x_v_gamma}~c) and in the controlled regime,
  Fig.~\ref{SinglePulse}~a), we see that the central $\gamma-$spike 
  is slightly larger while the both side spikes are significantly
  damped. This effect 
  becomes much more pronounced when we compare the filtered radiation plots,
  Fig.~\ref{spectrum_atto}~b) and Fig.~\ref{SinglePulse}~b). The control
  signal allows us to select the single attosecond pulse corresponding
  to the highest $\gamma-$spike in the surface motion.

  Varying the control parameters $a_0/a_c$, $\omega_0/\omega_c$ and
  $\Delta\phi$ we are able to select different attosecond pulses
  one-by-one or in groups out of the original pulse train.
 

  To recapitulate, we studied analytically and numerically the dynamics of
  the apparent reflection point at the overdense plasma surface. We
  have shown that the velocity of this point is a smooth 
  function of time. However, the corresponding $\gamma$-factor has
  quasi-singularities or spikes when the surface velocity approaches 
  the speed of light. These ultra-relativistic spikes are
  responsible for the high harmonic generation in the form of an 
  attosecond pulse train. We show that the attosecond pulse emission
  can be efficiently controlled by managing the laser
  polarization. This is done by adding a low intensity control pulse
  with perpendicular polarization and frequency slightly
  different from that of the driving pulse. This relativistic plasma control
  allows to gate a single attosecond pulse or a prescribed group of
  attosecond pulses. 

  This work has been supported in parts by DFG Transregio 18 and by DFG 
  Graduierten Kolleg 1203.


\end{document}